\documentstyle[]{article}
\newlength{\aivwidth}   
\newlength{\tmpwidth}   
\title{ Dynamics of overlapping vortices in complex scalar fields }
\author{Jacek Dziarmaga  \\
        Jagellonian University, Institute of Physics, \\
        Reymonta 4, 30-059 Krak\'ow, Poland
        \thanks{address from October 1, 1995:
        Department of Mathematical Sciences, University of Durham,
        South Road, Durham, DH1 3LE, United Kingdom;
        e-mail address: J.P.Dziarmaga@durham.ac.uk }}
\date{March 12, 1995 (Revised on November 2, 1995)}
\begin{document}
\maketitle
\begin{abstract}
We investigate dynamics of overlapping vortices
in the nonlinear Schr\"{o}dinger equation, the nonlinear heat equation
and in the equation with an intermediate Schr\"{o}dinger-diffusion dynamics.
Because of formal similarity on a perturbative level we discuss also the
nonlinear wave equation (Goldstone model). Special
solutions are found like vortex helices, double-helices and braids,
breather states and vortex mouths. A pair of vortices in the Goldstone
model scatters by the right angle in the head-on collision. It is found
that in a dissipative system there is a characteristic lenght scale above
which vortices can be entangled but below which the entanglement is unstable.
\end{abstract}
\newcommand{\be}{\begin{equation}\label}
\newcommand{\ee}{\end{equation}}
\newcommand{\ba}{\begin{eqnarray}\label}
\newcommand{\ea}{\end{eqnarray}}
\newcommand{\pl}{\partial}
\newcommand{\lb}{\lambda}
\newcommand{\kp}{\kappa}
\newcommand{\tr}{\triangle}
\newcommand{\dt}{\delta}
\newcommand{\bt}{\beta}
\newcommand{\al}{\alpha}
\newcommand{\gm}{\gamma}

\section{Introduction}

   The dynamics of the superfluid helium condensate at zero
temperature \cite{don} was proposed
to be described by the nonlinear Schr\"{o}dinger equation \cite{pit}.
Since then many approaches have been developed to study the
dynamics of vortices being topological excitations of the NLSE.
Lund \cite{lund} derives equations of motion for $3$-dimensional vortex
filaments with the help of the effective action method. The effective
model takes the form of an action at a distance theory. As anticipated
\cite{lund}, such a method can not describe many details of the dynamics.
Lee \cite{lee} rederives Lund's equations
but in addition takes into account sound radiation
from a moving vortex filament. His method originates from earlier
similar developements in relativistic field theories \cite{rel}.

   The dynamics of well separated vortices on a plane has been studied
by Neu \cite{neu}. The basic assumption of his method is that the modulus of
the scalar field around a given vortex is undisturbed by other distant
vortices. Interactions are mediated only by the phase of the complex scalar
field. Such a simplification is not reliable for vortices with
overlapping cores.

   In this paper we are going to consider the dynamics of vortices
when their cores strongly overlap. The method can be outlined as follows.
We take as a background configuration a planar vortex solution with
winding number say 2. Then we consider small fluctuations around this
background. The double vortex can be viewed as a superposition of two
unit vortices. As the double vortex has higher energy then two
single vortices, it is likely to split into vortices with winding
number 1. This splitting is described in our perturbative calculation
as a mode with dipole symmetry. Its energy is negative as it should be
for a splitting mode. One can also consider axially symmetric modes
which do not split the double vortex but instead they change its width.
They are usually called breathers or pulsons. Another possibility
is a mode with $Z_{3}$ symmetry. The double vortex might uniformly
split into 3 unit vortices leaving one antivortex at the center.
The perturbative calculation shows this way of splitting does not
take place for the double vortex. However for a vortex with winding
number 3 there is a $Z_{4}$-symmetric mode which describes
a decay of the triple vortex into 4 unit vortices with one antivortex
left at the origin. We generalize these considerations allowing
the modes to vary along the line of the background vortex. It this
setting the double vortex can be split into parallel unit vortices
and later on these vortices can be twisted to form a twisted pair
or a double helix. It turns out that there is a critical wave-lenght
at which such a double-helix is static.

   Our analysis is carried out not only for the nonlinear Schr\"{o}dinger
equation (NLSE) but also for a diffusive model we call after Neu
\cite{neu} the nonlinear heat equation (NLHE). We discuss
an intermediate Schr\"{o}dinger-diffusive dynamics case and that
of the nonlinear wave equation (NLWE) also known as the Goldstone model.
The physical reason to consider diffusive and hamiltonian dynamics
in one paper is that systems like superfluid helium change their
dynamical properties according to external conditions such as temperature.
At zero temperature the dynamics is hamiltonian but as we move towards
the phase transition it gradually becomes more diffusive. Thus the different
equations can describe one physical system but in different regimes.
There are also important mathematical reasons for the unified treatement.
First of all there is a set of static solutions, like static double helices,
which are common to all the considered equations. The eigenvalue problem
for time-dependent modes turns out to be formally the same for the NLWE
and the NLHE. Solving this problem we kill two birds with one stone.
The other reason is that there is a direct corespondence between
the modes in the NLHE case and those in the NLSE. Although the
at first sight temptating analytical continution in the complex time plane
fails, a large class of perturbative NLHE's solutions can be
mapped into corresponding perturbative solutions of the NLSE.
The analytical continuation does not apply to the modes themselves but it
does apply to their eigenvalues. Some NLHE's modes do not map to the
NLSE. However they still can be mapped to the intermediate diffusive
equation with some admixture of the Schr\"{o}dinger dynamics.

\section{Field equations and vortex solutions}

  We will consider three types of nonlinear field equations. The first
two are the nonlinear heat equation (NLHE)
\be{10}
\psi_{,t}=\tr\psi+(1-\mid\psi\mid^{2})\psi\equiv
-\frac{\dt F}{\dt\psi^{\star}} \;\;
\ee
and the nonlinear Schr\"{o}dinger equation (NLSE)
\be{20}
-i\psi_{,t}=\tr\psi+(1-\mid\psi\mid^{2})\psi\equiv
-\frac{\dt H}{\dt\psi^{\star}} \;\;.
\ee
where $\tr=\mbox{\boldmath $\nabla$}^{2}$ is a Laplacian.
The free energy and the Hamiltonian are
\be{30}
F=H=\int d^{3}x\;[\nabla\psi^{\star}\nabla\psi
                 +\frac{1}{2}(1-\mid\psi\mid)^{2}] \;\;.
\ee
The third equation is the nonlinear wave equation (NLWE)
\be{40}
\psi_{,tt}-\tr\psi-(1-\mid\psi\mid^{2})\psi\equiv
\frac{\dt L}{\dt\psi^{\star}}=0 \;\;,
\ee
with the Lagrangian
\be{50}
L=\int \;d^{3}x\;[\pl_{\mu}\psi^{\star}\pl^{\mu}\psi-
                    \frac{1}{2}(1-\mid\psi\mid^{2})^{2}] \;\;.
\ee
For time-independent fields all the three models reduce to the same
static equation
\be{60}
\tr\psi+(1-\mid\psi\mid^{2})\psi=0 \;\;.
\ee
It is well known that such an equation admits topological vortices.
A rotationally symmetric vortex solution can be obtained in the form
\be{70}
\tilde{\psi}(r,\theta)=f_{n}(r)e^{in\theta}\;\;,
\ee
where $(r,\theta)$ are polar coordinates. Substitution of this Ansatz
to Eq.(\ref{60}) yields
\be{80}
\tr_{n}f_{n}+(1-f_{n}^{2})f_{n}=0 \;\;,
\ee
where
$\tr_{n}\equiv
\frac{d^{2}}{dr^{2}}+\frac{1}{r}\frac{d}{dr}-\frac{n^{2}}{r^{2}}$.
The boundary condition at infinity is $f_{n}(\infty)=1$. The phase of the
complex scalar field is multivalued at the origin so its modulus
must vanish there, $f_{n}(0)=0$. If we restrict to regular solutions
the asymptotes at the origin and at infinity will turn out to be
\ba{90}
f_{n}(r)&\approx & f_{0}r^{n}+... \;\;,\;\; r\rightarrow 0 \;\;,\nonumber\\
f_{n}(r)&\approx & 1-\frac{n^{2}}{2r^{2}}+... \;\;,\;\; r\rightarrow\infty
\;\;. \ea
 From now on we will restrict to positive $n$.

\section{Fluctuations around vortex background in nonlinear heat and
                                            nonlinear wave equation}

 Let the background solution be a straight-linear vortex along
the z-axis with winding number n and the wavefunction
$\tilde{\psi}$, see (\ref{70}..\ref{90}).
The particle density of the background will be denoted by
$\rho=\tilde{\psi}^{\star}\tilde{\psi}$.
The equation satisfied by the field fluctuation $\dt\psi$ within
the NLHE is
\be{1.10}
-\dt\psi_{,t}=-\tr\dt\psi+(2\rho-1)\dt\psi
                       +\rho e^{2in\theta}\dt\psi^{\star} \;\;.
\ee
One could try decomposition of the solution into Fourier modes both
in $\theta$ and $z$, say $e^{ip\theta+ikz}$. However such a single Fourier
mode can not be a solution to Eq.(\ref{1.10}) because of the last term,
$\rho e^{2in\theta}\dt\psi^{\star}$, which couples Fourier modes into pairs.
The general solution can be obtained as a sum (or an integral) over
pairs of Fourier modes like
\be{1.20}
\xi(E,k,p) f_{0} e^{-Et}[
e^{ikz}e^{ip\theta}u_{1}(r)+
e^{-ikz}e^{i(2n-p)\theta}u_{2}(r)]
\ee
or
\be{1.25}
\bar{\xi}(E,k,p) f_{0} e^{-Et}e^{-kz}[
e^{ip\theta}u_{1}(r)+e^{i(2n-p)\theta}u_{2}(r)] \;\;,
\ee
where $\xi$'s are complex amplitudes of the modes and $f_{0}$ is a constant
from Eqs.(\ref{90}) which is introduced to provide the modes with a convenient
normalisation. The $z$-dependence factorizes out because the background
solution is $z$-independent. $p$'s are integers
which can range from $-\infty$ to $n$. Modes with $p>n$ are identical to
those with $p<n$ e.g. the $p=n+1$ modes can be identified with the $p=n-1$
modes. The eigenvalues $(E,k)$ can in principle take arbitrary
real values but only the modes with finite energy (or free energy) are
physical. The functions $u_{1},u_{2}$ in Eq.(\ref{1.20}) are solutions to
the eigenvalue problem
\ba{1.30}
&&(E-k^{2})u_{1}=-\tr_{p}u_{1}+(2\rho-1)u_{1}+\rho u_{2} \;\;,\nonumber\\
&&(E-k^{2})u_{2}=-\tr_{2n-p}u_{2}+(2\rho-1)u_{2}+\rho u_{1} \;\;,
\ea
where $\tr$'s are defined below Eq.(\ref{80}). The functions
$u_{1},u_{2}$ in Eq.(\ref{1.25}) are solutions of the same eigenvalue
problem as in Eqs.(\ref{1.30}) but with $k^{2}$ replaced by $-k^{2}$. From
now on we will allow $k^{2}$ to be a positive or negative real number.
$k^{2}>0$ will refer to the modes in Eq.(\ref{1.20}). $k^{2}<0$ will refer
to those in Eq.(\ref{1.25}) so that $k$ is to be replaced by $Im(k)$ there.

We can simplify Eqs.(\ref{1.30}) by  a replacement $E-k^{2}=\omega$,
\ba{1.40}
&&\omega u_{1}=-\tr_{p}u_{1}+(2\rho-1)u_{1}+\rho u_{2} \;\;,\nonumber\\
&&\omega u_{2}=-\tr_{2n-p}u_{2}+(2\rho-1)u_{2}+\rho u_{1} \;\;.
\ea
With equations (\ref{1.40}) at hand we are able to give a physical
interpretation to the modes with various $p$. If one restricts
to regular solutions one can easily find the asymptotes of the profile
functions close to the origin
\ba{1.50}
u_{1}(r)&\approx & u_{1}^{0}  r^{\mid p\mid}\;\;,\nonumber\\
u_{2}(r)&\approx & u_{2}^{0}  r^{\mid 2n-p\mid}+... \;\;,
\ea
where $u_{1}^{0},u_{2}^{0}$ are real constants which have to be choosen
so as to meet the conditions of regularity and fast convergence at infinity.
Eqs.(\ref{1.40}) are a set of linear homogenous differential equations so
the constants can be multiplied by a common factor. In other words
we can redefine the constants by an appropriate rescaling of the
overall amplitudes $\xi$ in Eqs.(\ref{1.20},\ref{1.25}). Thus, provided
that $u_{1}^{0}\neq 0$, we can choose $u_{1}^{0}=1$ and call the still free
$u_{2}^{0}=a$. In this case we are left with just one coefficient $a$
which we can vary to obtain solutions with acceptable asymptotes
at infinity. For $u_{1}^{0}=0$ the coefficient $u_{2}^{0}$ can be
rescaled to $1$, $u_{2}^{0}=1$. In this case, which is a set of measure
zero, there is no free parameter to remove singularities or slowly
convergent asymptotes at infinity and it turns out that there indeed is no
solution. Thus from now on we restrict our attention to the first case
\ba{1.55}
&&u_{1}(r)\approx r^{\mid p \mid}+... \;\;,\nonumber\\
&&u_{2}(r)\approx a r^{\mid 2n-p \mid}+... \;\;.
\ea
In the limit of very small $w=x+iy$ or close to the vortex axis,
for the modes in Eq.(\ref{1.20}), the total scalar field looks like
\be{1.60}
 \psi=\tilde{\psi}+\dt\psi \sim  w^{n}+\xi(E,k,p) e^{-Et+ikz} w^{p}
\ee
for $p\geq 0$.  The zeros of this polynomial coincide with the zeros
of the total scalar field.

   Let us consider the double vortex background,
$n=2$. For $p=0$ we have $\psi\sim w^{2}+\xi(E,k,0) e^{-Et+ikz}$. At $t=0$
and on the plane $z=0$ we have two zeros at complex roots of $-\xi$.
For $E>0$ these zeros will with time shrink down to $w=0$.
For nonzero $k$ the two lines of vanishing scalar field form a double helix.
If there is a solution for $k$ then there is also a solution for $-k$.
We can combine the two to obtain e.g. $\psi\sim w^{n}+\xi e^{-Et}\cos{kz}$
with $\xi$, say, real and positive constant. Let us fix time $t$ but
consider  what happens as we vary $z$. Close to $z=0$ the zeros lie on
$y$-axis. As we increase $z$ the zeros shrink to $w=0$ but then reappear
but rotated by the right angle. This kind of solution will be called
a double-braid.

   For general $n,p$ we get $n-p$ uniformly split vortices and $p$
vortices left at the origin. If $p<0$ there are $-p$ antivortices
at the origin. The case $n=p$ is exceptional. There is no splitting
but just a change of vortex width.

  Similar mode decomposition as for the NLHE (\ref{1.20},\ref{1.25}) can be
performed for the NLWE,
\be{1.65}
\xi(E,k,p) f_{0} [e^{iEt+ikz}u_{1}(r)+e^{-iEt-ikz}e^{i(2n-p)\theta}u_{2}(r)]
\;\;. \ee
Substitution of this Ansatz to the NLWE linearised in fluctuations
\be{1.90}
-\frac{\pl^{2}}{\pl t^{2}} \dt\psi=-\tr\dt\psi+(2\rho-1)\dt\psi
                       +\rho e^{2in\theta}\dt\psi^{\star} \;\;.
\ee
once again leads to Eqs.(\ref{1.40}) but this time
$\omega=E^{2}-k^{2}$. The Ansatz (\ref{1.65}) can be
generalised to admit $E^{2}$ or $k^{2}$ negative.
If say $E^{2}$ is negative one should replace the two time-dependent
exponents in the Ansatz (\ref{1.65}) by one exponent
$e^{-Im(E)t}$ standing in front of the mode. An analogous rearrangement
should be done if $k^{2}<0$.

  Since both the fluctuations in the NLHE and in the NLWE satisfy
Eqs.(\ref{1.40}) it is the highest time to look for their solutions.

\subsection{Asymptotes of the splitting modes at infinity}

    At infinity the asymptotic form of Eqs.(\ref{1.40}) is
\ba{1.100}
&& 0=\tr u_{1} + (\omega-1)u_{1} - u_{2}    \;\;,\nonumber\\
&& 0=\tr u_{2} - u_{1} + (\omega-1)u_{2}    \;\;.
\ea
These equations can be combined to give
\ba{1.110}
&& 0=\tr u_{+} + (\omega-2)u_{+}    \;\;,\nonumber\\
&& 0=\tr u_{-} + \omega u_{-}    \;\;,
\ea
where $u_{+}=u_{1}+u_{2}$ and $u_{-}=u_{1}-u_{2}$.

For $\omega >2$ both $u_{+}$ and $u_{-}$ approach combinations
of Bessel functions. In the range $0<\omega<2$, $u_{+}$'s
asymptote is a linear combination of the modified Bessel functions
while $u_{-}$ still
falls down like $1/\sqrt{r}$. Finally for $\omega<0$ both $u_{+}$
and $u_{-}$ are independent combinations of the modified Bessel functions.
The cases of $\omega=0,2$ need a separate treatement.

  We can accept only localised modes with finite energy or free energy.
With such a restriction we have to exclude both exponentially divergent
solutions and all the solutions with Bessel function-like asymptotes.
For $\omega=2$ there are slowly falling down oscillations
in the asymptote of $u_{-}$. Thus the range of $\omega$ where we
can look for acceptable solutions is restricted to $\omega\leq 0$.
In this range there are two divergent asymptotes at infinity.
For any given $\omega$ we can use the parameter $a$ in Eq.(\ref{1.55})
to fine tune one of the divergent asymptotes to zero. The second can be
removed by an appropriate choice of $\omega$. Such a distinguished $\omega$
is just the eigenvalue we are looking for.

\subsection{Bound states for $p=0..n-1$ }

There are analytical solutions for $p=n-1$ and $\omega=0$. They can be
constructed as $\dt\psi=\pl_{x}\tilde{\psi}(r,\theta)$, where
$\tilde{\psi}=f(r)e^{in\theta}$ is the background n-vortex solution. The
profile functions of such zero modes are
\ba{f.10}
&&u_{1}=\frac{1}{2}[f'(r)+\frac{nf(r)}{r}] \;\;\nonumber\\
&&u_{2}=\frac{1}{2}[f'(r)-\frac{nf(r)}{r}] \;\;.
\ea
The energy of these zero modes similarly as the energy
of the background vortex is logarythmically divergent.
They give rise to the following solution of the NLWE for small t
\be{f.20}
\dt\psi/f_{0}=-vt[u_{1}(r)+e^{2i\theta}u_{2}(r)] \;\;,
\ee
where $v$ is a complex velocity. For $n=1$ the zero of the scalar field
coincides with the zero of the polynomial $w-vt$. The solution is just
a perturbative approximation to a planar vortex moving with a constant
velocity. This solution becomes more interesting if we admit $k^{2}\neq
0$ to excite travelling waves with a dispersion
relation $E^{2}-k^{2}=0$. The line of vanishing scalar field coincides with
zeros of a polynomial $w(t,z)-F_{1}(t+z)-F_{2}(t-z)$, where $F$'s are arbitrary
functions. This solution is a perturbative approximation to Vachaspatis'
travelling waves \cite{vach}.

   For the solution with $\omega=0$ and $n=1,p=0$, the dispersion
relation in the NLHE case is $E-k^{2}=0$. The line of zero scalar field is
deformed from the $z$-axis, $w(t,z)=0$, to the helix
$w(t,z)=\xi(k^{2},k,0) \exp(-k^{2}t+ikz)$.
Any initial disturbance of this form (for example a helical standing
wave) will shrink down to the unperturbed straight-linear vortex.

  We have found some bound states for $\omega<0$.
Let me explain the method first. The asymptotes at infinity are
\ba{f.70}
&&u_{-}=A\frac{\exp(-\sqrt{\mid\omega\mid}\; r)}{\sqrt{r}}
     +B\frac{\exp(+\sqrt{\mid\omega\mid}\; r)}{\sqrt{r}} \;\;,\nonumber\\
&&u_{+}=C\frac{\exp[-\sqrt{\mid\omega\mid+2}\; r]}{\sqrt{r}}
     +D\frac{\exp[+\sqrt{\mid\omega\mid+2}\; r]}{\sqrt{r}} \;\;.
\ea
We want to find such $\omega$'s that B and D vanish simultaneously.
In Eqs.(\ref{1.55}) there is only one free parameter $a$.
For each $\omega$ we can choose such a value of $a$ that the
coefficient $D$ vanishes. Once $D$ is removed the coefficient $B$
remains in general nonzero. This procedure may be used to determine
the function $B(\omega)\equiv B(\omega\mid D=0)$. We assume the function
to be continuous. If there are two $\omega$'s with opposite signs of $B$,
there must be such a value of $\omega$ in between that $B=0$. This is the
way in which one can find a bound state solution to Eqs.(\ref{1.40}).

   For $n=1$ and $p=0$ we have scanned
a wide range of negative $\omega$'s finding the function $B$ to
be always positive. More fruitful was the search for $n=2,p=0$.
We have found a bound state for $\omega=-0.168\equiv -k_{0}^{2}$.
The static modes which arise from this state are the same
for all the considered models. The modes satisfy $k^{2}=k_{0}^{2}$.
The lines of zero scalar field are in general the roots of the complex
polynomial $w^{2}-\xi_{1}e^{ik_{0}z}-\xi_{2}e^{-ik_{0}z}$.
If only $\xi_{1}\neq 0$, the solution is a static double helix,
see the bottom part of Fig.1. Another characteristic solution is that with
$\xi_{1}=\xi_{2}$. This is a static double braid, see Fig.2.
The static double helices or braids exist only for the special
characteristic wave-lenght $L_{0}=2\pi/k_{0}$.
Perturbations with other wave-lenghts are no longer static.

   For the NLHE the dispersion relation is $E=k^{2}-k_{0}^{2}$.
$E$ is negative for fluctuations with wave-lenghts greater then $L_{0}$.
The external radius of such double-helices or braids is growing
with time like $\exp(\frac{1}{2}Et)$. Because of the string tension they
must stabilise at some larger radius which we are not able to reach by our
perturbative analysis. For short wave-lenght fluctuations
$E$ is positive and such double-helices or braids shrink down
like $\exp(-\frac{1}{2}Et)$ to the unperturbed straight-linear vortex
configuration.

   The NLWE has a different dispersion relation,
namely $E^{2}-k^{2}=-k_{0}^{2}$.
For generic values of positive $k^{2}$ and $E^{2}$ the double-helix at
the bottom of Fig.1 and the braid in Fig.2 move up or down the z-axis with
a phase velocity less then 1. Such modes should have been expected
since they are just Lorenz boosted static helices
or braids. More interesting are the modes with negative
$E^{2}$: $-E^{2}=k_{0}^{2}-k^{2}$. They exist for long wave-lenght
fluctuations as compared to the critical wave-lenght $L_{0}$. Such double
helices can decay or expand with a passage of time.
There is also a combination for which the external radius
evolves like $\sqrt{\sinh(\sqrt{-E^{2}}\;t)}$. This can describe
a double-helix (braid) shrinking to the basic solution and then
expanding but instantenously rotated around the z-axis by the right angle.

   In the limit of infinite wave-lenght
we obtain a pair of parallel vortices. Vortices
in the NLHE repel one another and the separation of their zeros
grows like $\exp(\frac{1}{2}k_{0}t)$. If they started right from the
unperturbed solution they would need infinite time to separate but
any generic initial dipole fluctuation should substantially
lower this time.
For a similar pair in the NLWE, the most interesting time evolution
is that in which the relative position evolves like
$\sqrt{\sinh(k_{0}t)}$. The positions of the zeros of the scalar field
are the same as the zeros of the complex polynomial $w^{2}-\sinh(k_{0}t)$.
For negative times vortices approach one another along the
y-axis, at the time t=0 they concide and for later times
they split but this time along the x-axis. This is the right angle
scattering in the head-on collision of two vortices. This result
can be generalised to symmetric collisions of n vortices to give
the $\frac{\pi}{n}$ scattering. A physical importance of this simple
right-angle scattering
has been realised by Manton \cite{manton}. Its net effect for multivortex
system's thermodynamics is that vortices behave as if they had
finite cores - there is a net excluded area for a planar vortex liquid
\cite{manton}.

\subsection{Bound states for $p=n$}

  We have seen, see Eq.(\ref{1.110}), that the asymptotic equations
for $u_{+}$ and $u_{-}$ decouple. For $p=n$ the two equations
decouple for any $r$. Eqs.(\ref{1.40}) can be rewritten as
\ba{l.10}
&&\omega u_{+}=-\tr_{n}u_{+}+(3\rho-1)u_{+}  \;\;,\nonumber\\
&&\omega u_{-}=-\tr_{n}u_{-}+(\rho-1)u_{-}  \;\;.
\ea
What are the bound states of these stationary Schr\"{o}dinger equations?
As $u_{-}$ is concerned, we have not found any
bound states for $n=1$ and $n=2$. The proof in Section 3.1 of
Ref.\cite{igor} applies here. According to this argument there are
no bound states for $u_{-}$.

The results for $u_{+}$ are more interesting. For n=1
there is one  bound state with energy
$\omega_{1}=1.806$. For n=2 the potential $(3\rho-1)$ is broader so that
the bound state energy is lower and amounts
to $\omega_{2}=1.613$. It seems that the energy of the ground state
is decreasing with increasing winding number. For sufficiently large
$n$ the next bound state is likely to appear. The existence of at least
one bound state for any $n$ can be proved following the lines
in Section 4 of Ref.\cite{igor}.

   The two modes look qualitatively the same for $n=1$ and $n=2$.
Once again let me begin with static modes. They satisfy $k^{2}=-\omega_{n}$.
The vortex exponentially broadens or tightens with z. An example is
shown in the top part of Fig.1. Such solutions may seem at first sight
to be simply diverging outside the perturbative regime and thus useless.
However looking at Fig.2 one can imagine that there is
a surface of the superfluid somewhere above and our mode is the way
in which the vortex mouth begins. Vortices prefer to be wider at
the surface then in the bulk.

   Now about the time-dependent solutions. In the NLHE the
dispersion relation is $E=\omega_{n}+k^{2}$. One can perturb the vortex
changing its core radius but the distortion will shrink down to the
unperturbed solution like $\exp{-Et}$. It shrinks the faster the shorter
is the wave-lenght of the fluctuation. In the NLWE, where the dispersion
relation is $E^{2}-k^{2}=\omega_{n}$, such a z-independent distortion
oscillates in time. Something like a "breather" or "pulson" state
forms. Making the last solution z-dependent, with $k^{2}>0$, we obtain
waves travelling along the vortex with a phase velocity
greater then 1.

   There are also solutions of Eqs.(\ref{l.10}) just on the verge
between bound states and scattering states. One of them is
$u_{-}=f(r)$ for $\omega=0$, where $f(r)$ is the moduli of the
background solution. Another solution is $u_{+}=f(\sqrt{3}r)$ for $\omega=2$.
These solutions would exist for any background, they have nothing
to do with the vortex. The first solution is a zero mode
due to a global gauge transformation
$\tilde{\psi}\rightarrow\tilde{\psi} e^{i\theta_{0}}$.
The second one would be $u_{+}=1$ for a uniform background $\rho=1$.
It is just an uniform massive oscillation of the scalar field's modulus
around its equilibrium value equal to 1.

\subsection{Bound states for $p<0$}

 As discussed in Section 3 such modes describe splitting of the
n-vortex into n unit vortices and in addition $\mid p\mid$
vortex-antivortex pairs. Such a decay is in principle possible
because the energy of n-vortex ($\sim n^{2}$) is higher then the
energy of n unit vortices ($\sim n$). The extra energy can be used
to create vortex-antivortex pairs. The maximal number of such
pairs has been estimated to be at best $\mid p\mid=n-1$, see \cite{h}.

  For n=1 and n=2 the result is negative. Double-vortex can be split
into 2 unit vortices but there is not enough energy to create a
$v-\bar{v}$ pair. Such a decay turns out to be possible for $n=3,p=-1$.
There is one antivortex left at the origin and 4 uniformly split
vortices. The corresponding eigenvalue is
$\omega=-0.103$. For the NLHE the mode is a diffusive splitting
with creation of one $v-\bar{v}$ pair. This way of splitting has
been observed in a direct numerical simulation, see the Figure 3 in
\cite{neu}. z-dependence can be introduced
giving rise to vortex 4-helix around a straight-linear antivortex or
4-braid with nodal points on the antivortex. For $k^{2}=-\omega$ the helices
and braids are static and for larger $k^{2}$ they are damped. For the NLWE
there are 4-helices and 4-braids at the critical wavelength. At other
wavelengths they are travelling waves. For parallel vortices the mode
describes head-on collision of 3 vortices on an antivortex center
resulting in $\frac{\pi}{n}$ scattering.

\section{Fluctuations in the nonlinear Schr\"{o}dinger equation regime}

   The discussion of the NLSE has been postponed until this section.
The NLSE can be obtained from the NLHE by a formal Wick rotation
$t=i\tau$
\be{s.10}
i\pl_{\tau}\psi=-\nabla^{2}\psi+(\psi^{\star}\psi-1)\psi \;\;.
\ee
However it does not mean we can do the same with solutions of
these equations. For example the eigenvalue problem for the fluctuations
around a vortex background in the NLSE
\be{s.15}
\xi(E,k,p) f_{0}[e^{iEt+ikz}e^{ip\theta}u_{1}(r)+
                          e^{-iEt-ikz}e^{i(2n-p)\theta}u_{2}(r)]
\ee
takes a slightly different form
\ba{s.20}
&&(-E+k^{2}+2\rho-1-\tr_{p})u_{1}+\rho u_{2}=0
                                                            \;\;,\nonumber\\
&&(+E+k^{2}+2\rho-1-\tr_{2n-p}) u_{2}+\rho u_{1}=0    \;\;
\ea
than in the NLHE.
Note that the equations differ by a sign in front of $E$. This makes
the eigenvalue problem different from that for the NLHE and NLWE,
compare with Eqs.(\ref{1.30}). This is where the analytical continuation in
the complex time plane fails on the perturbative level.

  Let us take a closer look at asymptotic properties of the solutions
to the eigenvalue problem (\ref{s.20}). The asymptotes at the origin
are the same as in the NLHE case, see Eq.(\ref{1.55}). As $E$ and $k^{2}$
can not be combined now into just one variable $\omega=E-k^{2}$, the
asymptotic behavior at infinity will depend on both $E$ and $k^{2}$
independently. In this limit the Eqs.(\ref{s.20}) become
\ba{s.30}
&&\tr u_{1} + (E-k^{2}-1)u_{1} - u_{2}=0  \;\;,\nonumber\\
&&\tr u_{2} - u_{1} + (-E-k^{2}-1)u_{2}=0  \;\;.
\ea
These equations can be diagonalized. The eigenvalues turn out to be
\be{s.40}
-(1+k^{2}) \stackrel{+}{-} \sqrt{ 1+E^{2} } \;\;.
\ee
For a positive eigenvalue we obtain a Bessel function-like asymptote
which does not converge fast enough to be acceptable. The two eigenvalues
are negative for $(E,k^{2})$ belonging to the area defined by
$k^{2}> -1 + \sqrt{ 1+E^{2} }$. In this "convergence area" there are
two exponentially decaying asymptotes and two exponentially divergent ones.
For any $(E,k^{2})$ one of the divergent asymptotes can be removed with
an appropriate choice of the constant $a$ in Eq.(\ref{1.55}). The other
one may happen to vanish along some lines in the $(E,k^{2})$-plane. These
distinguished lines are just the dispersion lines we are looking for.

  A straightforward but laborious way of solving the problem (\ref{s.20})
would be to scan the $(E,k^{2})$-plane in search of bound state solutions.
Let us check first what could we learn from analytical continuation of the
NLHE`s bound states to the NLSE.

  Let us consider the bound state of the NLHE
\be{s.50}
\bar{\dt\psi}=e^{-Et}[e^{ikz}e^{ip\theta}u_{1}(r)+
        e^{-ikz}e^{i(2n-p)\theta}u_{2}(r)] \;\;,
\ee
where the functions $u_{1}(r),u_{2}(r)$ satisfy Eqs.(\ref{1.30}). The
Wick rotated configuration takes the form
\be{s.60}
\bar{\dt\psi}=e^{-iEt}[e^{ikz}e^{ip\theta}u_{1}(r)+
        e^{-ikz}e^{i(2n-p)\theta}u_{2}(r)].
\ee
This configuration is a solution of the linearized NLSE but
only at $t=0$. Later on the fluctuation $\dt\psi$ must deviate
from the continued mode $\bar{\dt\psi}$ to $\dt\psi=\bar{\dt\psi}+\phi$.
We look for the deviation in the form
\be{s.70}
\phi=e^{ikz+ip\theta}[e^{iEt}w_{1}(r)+e^{-iEt}v_{1}(r)]+
     e^{-ikz+i(2n-p)\theta}[e^{-iEt}w_{2}(r)+e^{iEt}v_{2}(r)]  \;\;.
\ee
The profile functions $w$'s and $v$'s must satisfy the following sets of
inhomogenous differential equations
\ba{s.90}
&&(+E+k^{2}+2\rho-1-\tr_{p})w_{1}+\rho w_{2}=-2\rho u_{2}
                                                           \;\;,\nonumber\\
&&(-E+k^{2}+2\rho-1-\tr_{2n-p})w_{2}+\rho w_{1}=2\rho u_{1}    \;\;
\ea
and
\ba{s.100}
&&(-E+k^{2}+2\rho-1-\tr_{p})v_{1}+\rho v_{2}=2\rho u_{2}
                                                           \;\;,\nonumber\\
&&(+E+k^{2}+2\rho-1-\tr_{2n-p})v_{2}+\rho v_{1}=-2\rho u_{1}     \;\;.
\ea
The R.H.S.'s of the above inhomogenous equations are regular sources which
vanish exponentially at infinity. Close to $r=0$ the $w$'s
and $v$'s look like
\ba{s.110}
&& w_{1}(r)\approx w_{1}^{0} r^{\mid p\mid}+... \;\;,\\ \nonumber
&& w_{2}(r)\approx w_{2}^{0} r^{\mid 2n-p\mid}+... \;\;,\\ \nonumber
&& v_{1}(r)\approx v_{1}^{0} r^{\mid p\mid}+... \;\;,\\ \nonumber
&& v_{2}(r)\approx v_{2}^{0} r^{\mid 2n-p\mid}+... \;\;,
\ea
with $w^{0}$'s and $v^{0}$'s being constants which have to be choosen so as
to make the solutions convergent at infinity. This time the equations are
inhomogenous so this time the constants can not be rescaled to remove
a half of them. For each set of equations (\ref{s.90},\ref{s.100}) there
are two adjustable parameters, $(w_{1}^{0},w_{2}^{0})$ and
$(v_{1}^{0},v_{2}^{0})$ respectively.
As to the asymptotic properties at infinity, it is enough to note that
the differential operators in both sets (\ref{s.90},\ref{s.100}) are the
same as in the Eqs.(\ref{s.20}). If the eigenvalues $(E,k^{2})$ belong to
the convergence area $k^{2}>-1+\sqrt{1+E^{2}}$ than, for each set of
equations (\ref{s.90},\ref{s.100}), there are two divergent asymptotes at
infinity. These two asymptotes can be removed with the two free parameters
in the asymptotes close to the origin (\ref{s.110}).

   Thus it turns out that $\phi$ is a quickly convergent function provided
that $k^{2}>-1+\sqrt{1+E^{2}}$. For $(E,k^{2})$ belonging to this convergence
area the NLHE mode $\bar{\dt\psi}$ when analytically continued to the NLSE
gives rise to the mode $\dt\psi=\bar{\dt\psi}+\phi$. This means that
the parts of dispersion lines which belong to the convergence area
can be analytically continued from the NLHE to the NLSE. We have confirmed
this observation looking directly for the solutions of Eqs.(\ref{s.20}).
Let us then discuss the most characteristic solutions of the NLSE.

   For n=2 and p=0 the NLHE's dispersion relation is $E=k^{2}-k_{0}^{2}$.
The dispersion relation remains unchanged for the NLSE modes obtained
by mapping NLHE's solutions. For short wave-lenght double helices and braids
(Fig.1,Fig.2) the energy is positive. They can move up or down the z axis
as travelling waves. For $k^{2}=k^{2}_{0}$ they become static. In the range
of $k^{2}$ from $k^{2}_{0}$ to $\frac{1}{2}(1-\sqrt{1-4k^{2}_{0}})$ there are
once again travelling waves. Now the problem arises if we can continue the
dispersion
line $E=k^{2}-k_{0}^{2}$ ouside the convergence area, in particular to the
point
$(E=-k^{2}_{0},\; k^{2}=0)$ which corresponds to a pair of parallel
vortices rotating anticlockwise around their common center of mass
with angular velocity $\frac{1}{2}k^{2}_{0}$. Outside the convergence area
the long-range Bessel function-like asymptotes are unavoidable.
A well defined NLHE's mode when Wick rotated to the NLSE developes
a long range deviation $\phi$. Thus it seems that a rotating parallel
pair of vortices dissipates energy radiating sound waves. As the
potential between vortices is repulsive their mutual distance should be
growing with time.

   For the excitations $(n=1,p=0)$ of a single vortex the dispersion relation
is $E=k^{2}$. This line remains in the convergence area for all $k^{2}>0$.
The travelling waves on the single vortex move up or down the z-axis.
The phase velocity is falling down to zero as the wave-lenght tends to
infinity. These excitations can be identified with the experimentally
observed Kelvin modes \cite{don}.

   The dispersion relation for the breather states
$(p=n)$ is $E=\omega_{n}+k^{2}$ with $\omega_{n}$ positive but smaller
then $2$. For $n=1,2$ the whole dispersion line lies outside
the convergence area. It may happen that for some large enough $n$,
$\omega_{n}<1$. Then there would exist a finite part of the dispersion line
belonging to the convergence area. In any other case the breather modes
must radiate. For $k^{2}>0$ they travel up or down the vortex with some
phase velocity dependent on the wave-lenght. In the z-independent case
$k^{2}=0$ there is a breather state like in the NLWE - vortex width
is oscillating in time around the equilibrium value.  For $k^{2}<0$
there exists an exceptional point $(E=0,\; k^{2}=\omega_{n})$ where we
have a static vortex throat common to all the considered equations.

  In the $n=3,p=-1$ case 4 vortex lines uniformly split from the origin.
For $k^{2}$ large enough, the 4-helix or the 4-braid wave is travelling along
the central antivortex axis. The point on the dispersion line with $k^{2}=0$
lies outside the convergence area thus the mode with parallel vortices
rotating around the central antivortex is at best a radiative one.

   The fluctuations are similar to those in the NLWE. The difference is that
the dispersion relations for the modes in the NLSE are a nonrelativistic limit
of the dispersion relations for corresponding modes in the NLWE.
Some of the NLSE modes, if they exist at all, are unstable against decay by
radiation of sound waves.

\section{Mixed Schr\"{o}dinger-diffusive dynamics}

   So far we have considered only the NLHE and NLSE. In general we can replace
$t$ in the NLHE (\ref{1.10}) by $t\exp(i\gamma)$ to obtain the mixed
Schr\"{o}dinger-diffusion equation
\be{in.10}
-\psi_{,t}\cos\gamma+i\psi_{,t}\sin\gamma
         =-\tr\psi-(1-\mid\psi\mid^{2})\psi \;\;.
\ee
$\gamma$ can vary from $0$ to $\frac{\pi}{2}$. The fluctuation modes now take
the
form
\be{in.20}
\xi(E,k,p) f_{0}
e^{-Et\cos\gamma}
[e^{-iEt\sin\gamma}e^{ikz}e^{ip\theta}u_{1}(r)+
e^{+iEt\sin\gamma}e^{-ikz}e^{i(2n-p)\theta}u_{2}(r)],
\ee
where this time $u_{1},u_{2}$ are complex. To answer the question if the
exponentially localised NLHE's modes can be mapped into modes of the
intermediate
equation we have first to find its convergence area.

   The homogenous equations which $u_{1},u_{2}$ have to satisfy are
\ba{in.30}
&&[ \tr_{p}-k^{2}+(1-2\rho)+E] u_{1}-\rho u_{2}=0 \;\;,\nonumber \\
&&[ \tr_{2n-p}-k^{2}+(1-2\rho)+E \exp(-i\gamma)] u_{2}-\rho u_{1}=0 \;\;.
\ea
The real part of the eigenvalues of the matrix operator (for very large $r$) is
\be{in.40}
-(1+k^{2})+E\cos^{2}\gamma
              +Re\sqrt{1-E^{2}\sin^{2}\gamma e^{-2i\gamma}} \;\;.
\ee
The convergence area is a set of all such $(E,k^{2})$ that
\be{in.50}
k^{2}> -1+E\cos^{2}\gamma
              +\mid Re\sqrt{ 1-E^{2}\sin^{2}\gamma e^{-2i\gamma} } \mid \;\;,
\ee
where the real parts of both eigenvalues are negative. In this area,
by similar arguments as in the previous section, a deviation $\phi$ from
a continued NLHE's solution can be always made convergent.

   The dispersion relations are once again formally the same as for
the NLHE or the NLSE but as for the NLHE the sign of $E$ determines
whether a given mode is growing or decaying with time.

   A pair of parallel vortices not only rotates around the common
center of mass with the angular velocity $\;\frac{1}{2}k_{0}^{2}\sin\gamma\;$
but also their mutual distance is growing with time. The zeros of the scalar
field move along spiral lines. There is such a $\dt\gamma>0$ that for
$\gamma<\frac{\pi}{2}-\dt\gamma$ the dispersion relation $E=k^{2}-k_{0}^{2}$
for the $(n=2,p=0)$ mode can be continued from large positive $k^{2}$
down to the axis $k^{2}=0$ and a little bit below. Thus a pair of parallel
vortices rotating one around another is radiating sound waves in the NLSE
regime but a small dissipation makes them an exponentially localised solution.
The external radius of the long wave-lenght double-helix (Fig.1) or the
braid (Fig.2) is also growing with time until it stabilises thanks to
the string tension. If its wave-lenght is short, it prefers to shrink down to
the coincident two-vortex configuration.

  For single vortex  modes $(n=1,p=0)$ the dispersion relation
$E=k^{2}$ remains in the convergence area for all positive $k^{2}$.
A single-helix gradually shrinks down to the straight linear vortex.
Fluctuations in the vortex width, the same as in the NLSE case, remain outside
the convergence area.

\section{Summary}

   We have obtained a wide class of perturbative solutions like
helices, double-helices or braids. The question arises whether such
configurations really correspond to some exact solutions. We suppose
the answer to be yes. Configurations like helices (Kelvin waves)
or double helices were analysed within the models for widely
separated vortices (Kelvin waves are exact solutions to the NLWE \cite{vach}).
The braids can not be described by such models
because even if their amplitude is large vortices have to cross one another
at the nodal points. In the Bogomol'nyi limit of the Abelian Higgs
model some extra symmetries of the model enable an exact
construction of double-helices and braids \cite{helix}.

   The analysis was substantially simplified by various symmetries which
connect some of the considered equations. The perturbative calculations
in the NLWE and the NLHE regimes lead to the same stationary eigenvalue
problem. The perturbative solutions to the whole family of nonrelativistic
equations interpolating between the NLHE and the NLSE were constructed
from the NLHE's solutions. This construction shows that although the analytical
continuation of solutions in the complex time plain fails one can still
continue the dispersion relation, provided that dispersion line belongs
to the convergence area. The calculations in the NLHE were substantially
simpler then they would be in the corresponding problem for the NLSE.
The lesson is that it is convenient to map such problems to the NLHE
and later on continue the obtained NLHE's dispersion lines to the
equation under consideration.

\paragraph{Note added.}
A month after the first version of this paper the preprint by Goodband
and Hindmarsh \cite{h} appeared. The papers overlap in the
analysis on the NLWE.

\paragraph{Aknowledgement.}
I would like to thank Igor Barashenkov for drawing relevance of
Ref.\cite{igor} to my attention. This research was supported in part by the
KBN grant 2 P03B 085 08 and in part by the Foundation for Polish Science.

\section{Figure captions}

Fig.1. Superposition of two modes on a double vortex. The bottom
part is the double helix. In the top part the vortex mouth opens.
On the plotted surface the scalar field's modulus is equal to $\frac{1}{2}$.

Fig.2. The double braid. The double vortex splits into two unit vortex
cosine waves of the same polarisation. At each nodal point the polarisation
plane turns by the right angle. For n-braids the polarisation would turn by
the angle $\frac{\pi}{n}$ at each nodal point.


\begin{thebibliography}{99}
\bibitem{don} for a review see R.J.Donnelly,"Quantised vortices
              in helium II", Cambridge 1991,

\bibitem{pit} L.P.Pitaevskii, Z.Eksp.Teor.Fiz. 40 (1961) 646,
                                                      (JETP 13 (1961) 451),
\bibitem{lund} F.Lund, Phys.Lett.A 159 (1991) 245,

\bibitem{lee} K.Lee, preprint CU-TP-652 (cond-mat/9409046),

\bibitem{rel} U.Ben-Ya'acov, Nucl.Phys.B 382 (1992) 597, B 382 (1992) 616,
              J.Dziarmaga, Phys.Rev.D 48 (1993) 3809,

\bibitem{neu} J.C.Neu, Physica D 43 (1990) 385, D 43 (1990) 407,

\bibitem{manton} N.S.Manton, Nucl.Phys.B 400 (1993) 624,

\bibitem{igor} I.V.Barashenkov, A.D.Gocheva, V.G.Makhankov and I.V.Puzynin,
                                                  Physica D 34 (1989) 240,

\bibitem{vach} Vachaspati, T.Vachaspati, Phys.Lett.B 238 (1990) 41,

\bibitem{helix} J.Dziarmaga, Phys.Lett.B 328 (1994) 392.

\bibitem{h} M.Goodband and M.Hindmarsh, hep-ph/9503457.
\end{thebibliography}
\end{document}